\documentstyle[12pt]{article}
 
\input psfig

\newcommand{\nc}{\newcommand}

\nc{\eqr}[1]{(\ref{#1})}
\nc{\sref}[1]{\S\ref{#1}}
\nc{\tref}[1]{Table~\ref{#1}}
\nc{\fref}[1]{Figure~\ref{#1}}
\nc{\cref}[1]{Chapter~\ref{#1}}
\nc{\beq}{\begin{equation}}
\nc{\eeq}{\end{equation}}
\nc{\barray}{\begin{eqnarray}}
\nc{\earray}{\end{eqnarray}}
\nc{\barrayn}{\begin{eqnarray*}}
\nc{\earrayn}{\end{eqnarray*}}
\nc{\bcenter}{\begin{center}}
\nc{\ecenter}{\end{center}}
\nc{\lra}{\longrightarrow}
\nc{\ra}{\rightarrow}
\nc{\ADE}{$A$-$D$-$E$\ }
\nc{\setall}{\setcounter{equation}{0}
        \setcounter{definition}{0}
        \setcounter{lemma}{0}
        \setcounter{convention}{0}
        \setcounter{conjecture}{0}
        \setcounter{theorem}{0}
        \setcounter{proposition}{0}
        \setcounter{property}{0}
        \setcounter{fact}{0}
        \setcounter{corollary}{0}}
\nc{\setequation}{\setcounter{equation}{0}}
\nc{\hs}[1]{\hspace{#1 mm}}

\def\sCC{{\kern 0.27em\vrule height1.45ex width0.03em depth0em
	  \kern-0.30em\rm C}}
\def\C{{\mathchoice
  {\sCC}
  {\sCC}
  {\kern 0.225em \vrule height1.05ex width0.025em depth0em \kern-0.25em \rm C}
  {\kern 0.180em \vrule height0.78ex width0.02em depth0em \kern-0.2em \rm C}
	}}
\def\sHH{{\rm I\kern-.16em{}H}}
\def\H{{\mathchoice
  {\sHH}
  {\sHH}
  {\rm I\kern-.13em{}H}
  {\rm I\kern-.13em{}H} }}
\def\sNN{{\rm I\kern-.16em{}N}}
\def\N{{\mathchoice
  {\sNN}
  {\sNN}
  {\rm I\kern-.12em{}N}
  {\rm I\kern-.10em{}N} }}
\def\sPP{{\rm I\kern-.16em{}P}}
\def\P{{\mathchoice
  {\sPP}
  {\sPP}
  {\rm I\kern-.12em{}P}
  {\rm I\kern-.10em{}P} }}
\def\sQQ{{\kern 0.27em \vrule height1.45ex width0.03em depth0em
	  \kern-0.30em \rm Q}}
\def\Q{{\mathchoice
	{\sQQ}
	{\sQQ}
  {\kern 0.225em \vrule height1.05ex width0.025em depth0em \kern-0.25em \rm Q}
  {\kern 0.180em \vrule height0.78ex width0.020em depth0em \kern-0.20em \rm Q}
	}}
\def\sRR{{\rm I\kern-0.16em{}R}}
\def\R{{\mathchoice
  {\sRR}
  {\sRR}
  {\rm I\kern-0.12em{}R}
  {\rm I\kern-0.10em{}R} }}
\def\sZZ{{\rm Z\kern-0.32em{}Z}}
\def\Z{{\mathchoice
  {\sZZ}
  {\sZZ} 
  {\rm Z\kern-0.3em{}Z}     
  {\rm Z\kern-0.25em{}Z} }}  
\def\ZZZ{{\rm Z\kern-0.24em{}Z}}
\def\sKK{{\rm I\kern-0.16em{}K}}
\def\K{{\mathchoice
  {\sKK}
  {\sKK}
  {\rm I\kern-0.12em{}K}
  {\rm I\kern-0.10em{}K} }}
\def\cK{{\cal K}}
\def\ctg{\C^2/\Gamma}

\def\mt{\widetilde{M}}

\newtheorem{definition}{\bf DEFINITION}

\newtheorem{theorem}{\bf THEOREM}

\newtheorem{proposition}{\bf PROPOSITION}

\newtheorem{conjecture}{\bf CONJECTURE}

\renewcommand{\thefootnote}{\fnsymbol{footnote}}
\begin{document}

\begin{titlepage}
{\flushright{\small MIT-CTP-2837\\ hep-th/9903056\\}}
\begin{center}

{\LARGE Of McKay Correspondence,}\\
\vspace{3mm}
{\LARGE Non-linear Sigma-model}\\
\vspace{3mm}
{\LARGE and Conformal Field Theory}\\
\end{center}

\vspace{1cm}
\begin{center}
{\large Yang-Hui He\footnote{E-mail: yhe@ctp.mit.edu}}\\
\vspace{1mm}
{\it and}\\
\vspace{1mm}
{\large Jun S. Song\footnote{E-mail:
jssong@ctp.mit.edu.\\ Research supported in part
by the NSF and the U.S. Department of Energy under cooperative research
agreement $\#$DE-FC02-94ER40818.
}}\\
\vspace{.8cm}
{\it Center for Theoretical Physics}\\
{\it Massachusetts Institute of Technology}\\
{\it Cambridge, Massachusetts 02139}
\end{center}

\begin{abstract}
The ubiquitous \ADE classification has induced many proposals of 
often mysterious correspondences both in mathematics and physics.
The mathematics side includes quiver theory and
the McKay Correspondence which relates finite group
representation theory to Lie algebras as well as
crepant resolutions of Gorenstein singularities.  On the
physics side, we have the graph-theoretic
classification of the modular invariants of WZW models,
as well as the relation between the string theory nonlinear
$\sigma$-models 
and Landau-Ginzburg orbifolds.
We here propose a unification scheme which naturally incorporates all 
these correspondences of the \ADE type in two complex dimensions.
An intricate web of inter-relations is constructed, providing a
possible 
guideline to establish new directions of research 
or alternate pathways to the standing problems 
in higher dimensions.
\end{abstract}
\end{titlepage}
\renewcommand{\thefootnote}{\arabic{footnote}}


\section{Introduction}
\begin{figure}
\centerline{\psfig{figure=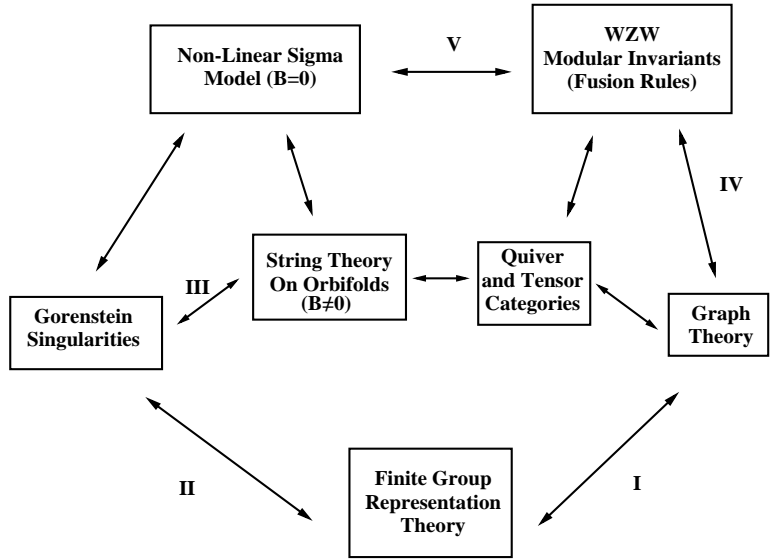,width=4.0in}}
\caption{The Myriad of Correspondences: it is the purpose of this
	paper 
	to elucidate these inter-relations in 2-dimensions, so as to
	motivate a similar coherent picture in higher dimensions.
	Most of the subsectors in this picture have been studied
	separately by mathematicians and physicists, but they are in
	fact not as disparate as they are guised.}
\label{fig:mother}
\end{figure}

This paper reviews the known facts about the various \ADE
classifications that arise in mathematics and string theory and
organizes them into a unified picture.  This picture serves as a guide
for our on-going work in higher dimensions and naturally incorporates 
diverse concepts in mathematics.

In the course of their research on supersymmetric Yang-Mills theories 
resulting from the type IIB D-branes on orbifold singularities \cite{he}, as
prompted by collective works in constructing (conformal) gauge theories
(e.g., \cite{Physics} \cite{Douglas} \cite{Kapustin} and references
therein), 
it was conjectured by Hanany and He that there may exist a McKay-type
correspondence between the bifundamental matter content and the modular 
invariant partition functions of the Wess-Zumino-Witten (WZW)
conformal field theory.  Phrased in another way, the correspondence,
if true, would
relate the Clebsch-Gordan  coefficients for tensor products of the 
irreducible representations
of finite subgroups of $SU(n)$ 
with the integrable characters for the affine algebras $\widehat{SU(n)}_k$
of some integral weight $k$.

Such a  relation has been well-studied in the case of $n=2$ and it
falls into 
an \ADE classification scheme \cite{CFT,GW,Gannon2}.
Evidences for what might occur
in the case of $n=3$ were presented in \cite{he} by computing the
Clebsch-Gordan coefficients extensively for  
the subgroups of $SU(3)$. Indications from the lattice
integrable model perspective were given in \cite{DiFrancesco}.

The natural question to pose is why there should be such correspondences.
Indeed, why should there be such an intricate chain of connections
among 
string theory on orbifolds, finite representation theory, graph
theory, affine 
characters and WZW modular invariants? In this paper, we hope to 
propose a unified quest
to answer this question from the point of view of the conformal field
theory description of Gorenstein singularities.  We also observe that
category theory seems to prove a common basis for all these theories.

We begin in two dimensions, where there have been numerous independent 
works in the past few decades in both mathematics and physics to
establish 
various correspondences.
In this case, the all-permeating theme is the \ADE classification.
In particular, there is the original McKay's correspondence between 
finite subgroups of $SU(2)$ and the \ADE Dynkin diagrams \cite{McKay} 
to which we henceforth refer as the {\it Algebraic McKay Correspondence}.
On the geometry side, the representation rings of these groups 
were related to the Groethendieck (cohomology) rings of the resolved
manifolds 
constructed from the Gorenstein singularity of the respective groups
\cite{threedim}; we shall refer to this as the {\it Geometric McKay
Correspondence}. 
Now from physics, studies in conformal field theory (CFT) have prompted many
beautiful connections among graph theory, fusion algebra, and modular 
invariants \cite{CFT,GW,Gannon2,minimal,gepner}. The classification
of the modular invariant partition function associated with 
$\widehat{SU(2)}$ Wess-Zumino-Witten (WZW) models also mysteriously 
falls into an \ADE type \cite{CIZ}. There have been some
recent 
attempts to explain this seeming accident from the
supersymmetric field theory and brane configurations perspective 
\cite{SD,ABKS}. In this paper we push from the direction of the 
Geometric McKay Correspondence and see how Calabi-Yau (CY) non-linear
sigma models 
constructed on the Gorenstein singularities associated with the finite
groups may be related to Kazama-Suzuki coset models 
\cite{minimal,gepner,vafa1,vafa2,witten,witten2}, which in turn can
 be related to 
the WZW models. This link would provide a natural setting for the emergence of
the \ADE classification of the modular invariants.
In due course, we will review and establish a catalog of 
inter-relations, whereby 
forming a complex web and unifying many independently noted
correspondences.  Moreover, we find a common theme of categorical
axioms  that all of these
theories seem to satisfy and suggest why the \ADE classification and its
extensions arise so naturally.
This web, presented in Figure~\ref{fig:mother}, is the central idea of
our paper.  Most of the correspondences in \fref{fig:mother} actually
have been discussed in the string theory literature although not all
at once in a unified manner with a mathematical tint.

Our purpose is two-fold.  Firstly, we shall show that tracing 
through the arrows in \fref{fig:mother}
 may help to enlighten the links that may seem
accidental. Moreover, and perhaps more importantly, we propose that
this program may be extended beyond two dimensions and hence 
beyond $A$-$D$-$E$. Indeed, algebraic geometers have done extensive research 
in generalizing
McKay's original correspondence to Gorenstein singularities of dimension
greater than 2 (\cite{Gonzales} to \cite{Brylinski}); many standing
conjectures exist in this respect. On the other hand, there is the
conjecture mentioned above regarding the $\widehat{SU(n)}_k$ WZW
 and the subgroups of $SU(n)$ in general.
It is our hope that Figure~\ref{fig:mother}
remains valid for $n > 2$ so that these conjectures may be attacked
by the new pathways we propose. We require and sincerely hope for the 
collaborative effort of many experts in mathematics and physics 
who may take interest in this attempt to
unify these various connections.

The outline of the paper follows the arrows drawn in
Figure~\ref{fig:mother}.  
We begin in \sref{sec:ubiquity} by
summarizing the ubiquitous \ADE classifications, and
\sref{sec:arrows} will be devoted to clarifying these \ADE links, while bearing
in mind how such ubiquity may permeate to
higher dimensions. It will be divided in to the following subsections:

\begin{itemize}
\item I. The link between representation
	 theory for finite groups and quiver graph theories (Algebraic
	 McKay); 
\item II. The link between finite groups and crepant 
	resolutions of Gorenstein singularities (Geometric McKay);
\item III. The link between resolved Gorenstein singularities, 
	Calabi-Yau manifolds and chiral rings for the associated
	non-linear sigma model 
        (Stringy Gorenstein resolution);
\item IV. The link between quiver graph theory for finite groups and WZW
	modular invariants in CFT, as discovered in the  study of
	of string orbifold theory (Conjecture in \cite{he});
\end{itemize}
	and finally, to complete the cycle of correspondences,
\begin{itemize}
\item V. The link between the singular geometry and its conformal
	field theory description provided by an orbifoldized coset
	construction which contains the  WZW theory.
\end{itemize}

\vspace{2mm}
\noindent
In \sref{sec:CFT} we discuss arrow V which
fills the gap between mathematics and physics, explaining why WZW
models have the magical properties that are so closely related to the
discrete subgroups of the unitary groups as well as to geometry.
From all these links arises \sref{sec:conj} 
which consists of our conjecture that there exists a conformal
field theory description of the Gorenstein singularities for higher
 dimensions, encoding the relevant information about the discrete
 groups and the cohomology ring.
In \sref{sec:ribbons}, we hint at how these vastly different fields may have
similar structures by appealing to the so-called ribbon and quiver categories.

Finally in the concluding remarks of \sref{sec:conclusion},
 we discuss the projection for future labors.

We here transcribe our observations with the hope 
they would spark a renewed interest in the study of McKay 
correspondence under a possibly new light of CFT, and vice versa.
We hope that Figure~\ref{fig:mother} will open up many
interesting and exciting pathways of research and that not only some
existing 
conjectures may be solved by new methods of attack, but also further
beautiful 
observations could be made.

\vspace{1cm}
\noindent
{\large\bf  Notations and Nomenclatures}\\
\noindent
We put a \ $\widetilde{}$\ \/ over a singular variety
to denote its resolved geometry. By dimension we mean
complex dimension unless stated otherwise.
Also by ``representation ring of $\Gamma$,'' we mean the 
ring formed by the tensor product decompositions of the irreducible 
representations of $\Gamma$.  The capital Roman numerals, I--V, in front
of the section headings correspond to the arrows in
Figure~\ref{fig:mother}.


\setall
\section{Ubiquity of \ADE Classifications} \label{sec:ubiquity}
\begin{figure}[ht]
\centerline{\psfig{figure=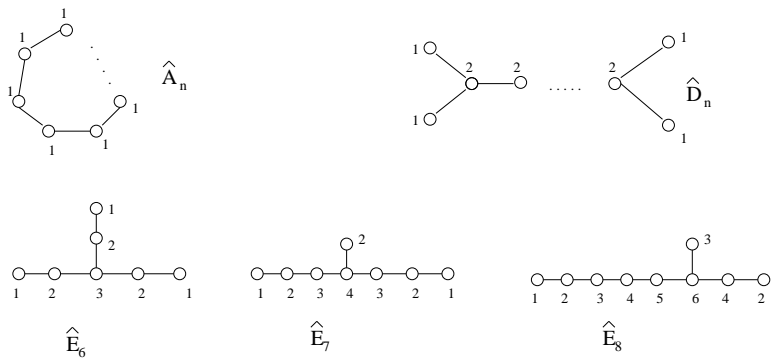,width=4.0in}}
\caption{The Affine Dynkin Diagrams and Labels.}
\end{figure}

\begin{table}[ht]
\vspace{5mm}
\begin{center}
\begin{tabular}{|p{1cm}|p{3cm}|p{4cm}|p{3cm}|}
\hline
 & Theory & Nodes & Matrices \\ 
\hline \hline
(a)& Finite Subgroup $\Gamma$ of $SU(2)$ & 
        Irreducible Representations &
        Clebsch-Gordan Coefficients\\
\hline
(b)& Simple Lie algebra of type $ADE$ & Simple Roots & 
        Extended Cartan matrix \\
\hline
(c)& Quiver Dynkin Diagrams & Dynkin Labels &
        Adjacency Matrix \\
\hline
(d)& Minimal Resolution $X\to\C^2/\Gamma$ &
        Irreducible Components of the Exceptional Divisor
        (Basis of $H_2(X,\Z)$) &
        Intersection Matrix \\
\hline
(e)& $\widehat{SU(2)_k}$ WZW Model &
        Modular Invariants / WZW Primary Operators &
        Fusion Coefficients \\
\hline
(f)& Landau-Ginzburg &
        Chiral Primary Operators & 
        Chiral Ring Coefficients \\
\hline
(g)& CY Nonlinear Sigma Model &
        Twisted Fields &
        Correlation Functions \\
\hline 
\end{tabular} 
\caption{\ADE Correspondences in 2-dimensions. The same
	graphs and their affine extensions appear in
	different theories.}\label{table:ADE}
\end{center}
\end{table}

In this section, we summarize the appearance of the \ADE
classifications in physics and mathematics and their commonalities.

It is now well-known that the complexity of particular algebraic and
geometric 
structures can often be organized into classification schemes of the 
\ADE type. The first hint of this structure  began in the 1884 work
of F. Klein in which he classified the discrete subgroups $\Gamma$
of $SU(2)$ \cite{Klein}.
These were noted to be in 1-1 correspondence with the Platonic solids in 
$\R^3$, and with some foresight, we write them as:
\begin{itemize}
\item type A: the cyclic groups (the regular polygons);
\item type D: the binary dihedral groups (the regular dihedrons) and 
\item type E: the binary tetrahedral (the tetrahedron), octahedral
	(the cube and the octahedron) and icosahedral (the
	dodecahedron and 
	the icosahedron) groups,
\end{itemize}
where we have placed in parenthesis next to each group the geometrical shape
for which it is the double cover of the symmetry group.

The ubiquity of Klein's original hint has persisted till the present day.
The \ADE scheme has manifested itself in such diverse fields as
representation theory of finite groups, quiver graph theory, Lie algebra
theory, algebraic geometry, conformal field theory, and string theory.
It will be the intent of the next section to explain the details of the
correspondences appearing in Table~\ref{table:ADE}, and we will subsequently
propose their extensions in the remainder of the paper.
%

\section{The Arrows of Figure 1.} \label{sec:arrows}
In this section, we explain the arrows appearing in Figure 1.  We verify
that there are compelling evidences in favor of the picture for 
the case of $2_\C$-dimensions, and we will propose its generalization to
higher dimensions in the
subsequent sections, hoping that it will lead to new insights on the
McKay correspondence as well as conformal field theory.

\subsection{(I) The Algebraic McKay Correspondence} \label{sec:alg-mckay}
In the full spirit of the omnipresent \ADE classification, it has been
noticed in 1980 by J. McKay that there exists a remarkable correspondence
between the discrete subgroups of $SU(2)$ and the affine Dynkin graphs
\cite{McKay}. 
Indeed, this is why we have labeled the subgroups in the
manner we have done.
\begin{definition}
For a finite group $\Gamma$, let $\left\{ r_i \right\}$ be its set of
irreducible representations (irreps), then we define the coefficients
$m_{ij}^k$ appearing in
\begin{equation}
r_k \otimes r_i = \bigoplus\limits_{j} m_{ij}^k r_j
\label{CG}
\end{equation}
to be the {\bf Clebsch-Gordan coefficients} of $\Gamma$.
\end{definition}
For $\Gamma \subset SU(2)$ McKay chose a fixed 
(not necessarily irreducible) representation
$R$ in lieu of general $k$ in \ref{CG} and defined matrices $m_{ij}^R$.
He has noted that up to automorphism, there always exists a
unique 2-dimensional representation,
which for type A is the tensor sum of 2 dual 1-dimensional irreps
and for all others the self-conjugate 2-dimensional irrep.
It is this $R=2$ which we choose and simply write the matrix as $m_{ij}$.
The remarkable observation of McKay can be summarized in the
following theorem;
the original proof was on a case-to-case basis and Steinberg gave a
unified proof in 1981 \cite{McKay}.
\begin{theorem}[{\sc McKay-Steinberg}]
For $\Gamma = A,D,E$, the matrix 
$m_{ij}$ is $2 I$ minus
the Cartan matrix of the 
affine extensions of the respective simply-laced simple Dynkin diagrams 
$\widehat{A}, \widehat{D}$ and $\widehat{E}$,
treated as undirected $C_2$-graphs (i.e., maximal eigenvalue of the
adjacency matrix is 2).
\end{theorem}
Moreover, the Dynkin labels of the nodes of the affine
Dynkin diagrams are
precisely the dimensions of the irreps.  Given a discrete subgroup
$\Gamma \subset SU(2)$, 
there thus exists a Dynkin diagram that
encodes the essential information about the representation ring of $\Gamma$.
Indeed the
number of nodes should equal to the number of irreps and thus by a
rudimentary 
fact in finite representation theory, subsequently equals the number of 
conjugacy classes of $\Gamma$. Furthermore, if we remove the node
corresponding to the trivial 1-dimensional (principal) representation,
we obtain the regular \ADE Dynkin diagrams. We present these facts in the
following diagram:
\vspace{3mm}
\bcenter
\parbox{1.2in}{Clebsch-Gordan Coefficients for $\Gamma=A,D,E$} \hs{2} 
$\longleftrightarrow$\hs{2} 
\parbox{1.2in}{Dynkin Diagram of
$\widehat{A},\widehat{D},\widehat{E}$}\hs{2}  
$\longleftrightarrow$\hs{2} \parbox{1.5in}{Cartan matrix and dual
Coxeter labels of  
	$\widehat{A},\widehat{D},\widehat{E}$}
\ecenter
\vspace{3mm}
This is Arrow I of Figure~\ref{fig:mother}.

Proofs and extension of McKay's results from geometric perspectives 
of this originally combinatorial/graph-theoretic 
theorem soon followed; they caused fervent activities in both algebraic
geometry and string theory (see e.g., \cite{orbifold,Gonzales,threedim}).
Let us first turn to the former.

\subsection{(II) The Geometric McKay Correspondence}
In this section, we are interested in crepant resolutions of Gorenstein
quotient singularities. 
\begin{definition}
The singularities of
$\C^n/\Gamma$ for $\Gamma\subset GL(n,\C)$ are called {\bf Gorenstein}
if there exists a nowhere-vanishing
holomorphic $n$-form\footnote{Gorenstein singularities thus provide
local 
models of singularities on Calabi-Yau manifolds.} 
on regular points.   
\end{definition}
Restricting $\Gamma$ to
$SU(n)$ would guarantee that the quotient singularities are
Gorenstein.  
\begin{definition}
We say that a smooth variety
$\widetilde{M}$ is a {\bf crepant} resolution of a singular
variety
$M$ if there exists a birational morphism $\pi:\widetilde{M}\ra M$
such that the canonical sheaves $\cK_M$ and $\cK_{\widetilde{M}}$ are
the same, 
or more precisely, if $\pi^* (\cK_M) = \cK_{\widetilde{M}}$.
\end{definition}

For $n\leq 3$, Gorenstein singularities always admit crepant
resolutions \cite{threedim}. On the other hand, 
in dimensions greater than 3, there
are known examples of terminal Gorenstein singularities which do not
admit crepant resolutions.  It is believed, however, that when the
order of $\Gamma$ is sufficiently larger than $n$, there exist crepant
resolutions for most of the groups.

The traditional \ADE classification is relevant in studying the
discrete subgroups of $SU(2)$ and resolutions of Gorenstein
singularities in two complex-dimensions.
Since we can choose an invariant Hermitian metric on $\C^2$, 
finite subgroups of $GL(2,\C )$ and $SL(2,\C)$ are conjugate to finite
subgroups of $U(2)$ and $SU(2)$, respectively.  Here,
motivated by the string compactification on manifolds of trivial
canonical bundle, we consider the linear
actions of non-trivial discrete subgroups $\Gamma$ of $SU(2)$ on $\C^2$.
Such quotient spaces $M=\ctg$, called {\it orbifolds},  have
fixed points which are
isolated Gorenstein singularities of the \ADE type studied by Felix Klein.  

As discussed in the previous sub-section, McKay\cite{McKay} has observed
a 1-1 correspondence between the non-identity conjugacy 
classes of discrete subgroups of  
$SU(2)$ and the Dynkin diagrams
of $A$-$D$-$E$ simply-laced Lie algebras, and this relation
in turn provides an
indirect correspondence between the orbifold singularities of
 $M$ and the Dynkin
diagrams. 
In fact, there exists a direct geometric correspondence between
the crepant resolutions of $M$ and the Dynkin diagrams. 
 Classical theorems in algebraic geometry tell us
that there exists a unique crepant resolution $(\widetilde{M},\pi)$ of the
Gorenstein
singularity of $M$ for all
$\Gamma\subset SU(2)$.  Furthermore, the exceptional divisor
$E=\pi^{-1} (0)$ is a 
compact, connected union of irreducible $1_{\C}$-dimensional curves of
genus zero\footnote{We will refer to them as $\P^1$ blow-ups.} such that
their intersection matrix is represented by 
the simply-laced Dynkin diagram
associated to $\Gamma$.  More precisely,  each node of the diagram
corresponds to an irreducible $\P^1$, and the  intersection matrix
is negative of 
the Cartan matrix of the Dynkin diagram such that two $\P^1$'s intersect
transversely at one point if and only if the two nodes are connected
by a line in the diagram.
In particular, we see that
the curves have self-intersection numbers $-2$ which exhibits the
singular nature of the orbifold upon blowing them down.
Simple consideration shows that these curves form a basis of the 
homology group $H_2(\widetilde{M},\Z )$ which is seen to coincide with
the root 
lattice of the associated Dynkin diagram by the above identification.
Now, combined with the algebraic McKay correspondence, this crepant
resolution  
picture  yields a 1-1 correspondence
between the basis of $H_2(\widetilde{M},\Z )$ and the non-identity
conjugacy classes of $\Gamma$.
We recapitulate the above discussion in the following diagram:
\vspace{3mm}
\bcenter
\parbox{1.2in}{$H_2(\widetilde{M},\Z)$ of the blow-up} \hs{2} 
$\longleftrightarrow$\hs{2} 
\parbox{1in}{Dynkin Diagram of $\Gamma$}\hs{2} 
$\longleftrightarrow$\hs{2} \parbox{1.5in}{Non-identity Conjugacy
Classes of $\Gamma$}
\ecenter
\vspace{3mm}
This is Arrow II in Figure~\ref{fig:mother}.
Note incidentally that one can think of irreducible representations as
being dual to 
conjugacy classes and hence as basis of $H^2(\widetilde{M},\Z)$.  This
poses 
a subtle question of which correspondence is more natural, but we will
ignore such issues in our discussions.

It turns out that $M$ is not only analytic but also
algebraic; that is, $M$ is isomorphic to $f^{-1} (0)$, where
$f:\C^3\rightarrow\C$
is one of the polynomials in \tref{rational} depending on $\Gamma$.
The orbifolds defined by the zero-loci of the polynomials are 
commonly referred to as the singular ALE spaces.

\begin{table}
\begin{center}
\begin{tabular}{||l|l|l||}  \hline
$f(x,y,z)$ & Subgroup $\Gamma$ & Order of $\Gamma$\\ \hline
$x^2 + y^2 +z^{k+1}$ & $A_k$ Cyclic & $k+1$\\ \hline
$x^2 + y^2z +z^{k-1}$ & $D_k$ Binary Dihedral & $4(k-2)$\\ \hline
$x^2 + y^4 +z^3$ & $E_6$ Binary Tetrahedral & 24 \\ \hline
$x^2 + y^3z +z^3$  & $E_7$ Binary Octahedral & 48\\ \hline
$x^2 + y^5 +z^3$ & $E_8$ Binary Icosahedral & 120\\ \hline
\end{tabular}
\caption{Algebraic Surfaces with Quotient Singularities \label{rational}}
\end{center}
\end{table}

\subsection{(II, III) McKay Correspondence and SCFT}

One of the first relevance of \ADE series in conformal field
theory appeared in attempts to classify $N=2$ superconformal
field theories (SCFT) with central charge $c<3$ \cite{minimal}.  
Furthermore,  the exact forms of the \ADE polynomials in
\tref{rational} appeared in a similar attempt to classify certain
classes of $N=2$ SCFT in terms of Landau-Ginzburg (LG) models.  The LG
super-potentials were precisely classified by the polynomials, and the
chiral ring and quantum numbers were computed with applications of
singularity theory \cite{vafa2}.  The LG theories which realize coset
models would appear again in this paper to link the WZW to geometry.

In this subsection, we review how string theory, when the $B$-field is
non-vanishing, resolves the orbifold 
singularity and how it encodes the information about the cohomology of 
the resolved manifold.  
Subsequently, we will consider the singular limit
of the conformal field theory on orbifolds by turning off the $B$-field, and
we will argue  that, in this singular limit, 
the $\widehat{SU(2)}_k$ WZW fusion ring 
inherits the information about the cohomology ring from the smooth theory.

\subsubsection{Orbifold Resolutions and Cohomology Classes} 
\label{sec:orbifold}
Our discussion here will be general and not restricted to $n=2$.
Many remarkable features of string theory stem from the fact that we can
``pull-back'' much of the physics on the target space to the
world-sheet, and as a result, the resulting
world-sheet conformal field theory somehow encodes the geometry of the
target space.  One example is that CFT is often\footnote{Not all CFT on
singular geometry are smooth.  For example, there are
examples of singular CFT's defined on singular backgrounds, such as
in the case of gauge symmetry enhancement of the Type IIA string theory
compactified on singular $K3$ where the $B$-field vanishes
\cite{aspinwall}.  Later, we will discuss a tensored coset model 
\cite{Ooguri-Vafa} 
describing this singular non-linear sigma model and relate it to the
algebraic McKay Correspondence.} insensitive to 
Gorenstein singularities and quantum effects revolve the singularity
so that the CFT is smooth.  More precisely, Aspinwall \cite{aspinwall2}
has shown that non-vanishing of the $NS$-$NS$ $B$-field makes the CFT
smooth.   In fact, string theory predicts the Euler characteristic
of the {\it resolved}\/ orbifold \cite{orbifold}; the local form of
the statement is

\begin{conjecture}[{\sc Stringy Euler characteristic}]\label{euler}  Let
$M=\C^n/\Gamma$ for $\Gamma\subset SU(n)$ a finite subgroup.  Then,
there exists a crepant resolution $\pi: \mt \ra M$ such that
\beq
	\chi (\mt) = | \{ \mbox{Conjugacy Classes of $\Gamma$}\}| \ .
\eeq
\end{conjecture}
Furthermore, the Hodge numbers of resolved orbifolds
were also predicted by Vafa for
CY manifolds realized as hypersurfaces in weighted projective
spaces  and by Zaslow for K\"{a}hler manifolds \cite{vafa4}.
In dimension three, it has been proved \cite{threedim,ito-reid} that
every  Gorenstein singularity
 admits a crepant resolution\footnote{In fact, a given Gorenstein
singularity generally admits many crepant resolutions \cite{Joyce}.
String theory so far has yielded two distinguished
desingularizations: the traditional CFT resolution without discrete
torsion and deformation with discrete torsion \cite{torsion}.  In this
paper, we are concerned only with K\"{a}hler resolutions without
discrete torsion.}
and that every
crepant resolution satisfies the
Conjecture~\ref{euler} and the Vafa-Zaslow Hodge number formulae.  For
higher dimensions, there are compelling evidences that the formulae are
satisfied by all crepant resolutions, when they exist.

As the Euler Characteristic in mathematics is naturally defined by the
Hodge numbers of cohomology classes, motivated by the works of string
theorists and the fact that $\mt$ has no odd-dimensional
cohomology\footnote{See \cite{ito-reid} for a discussion on this point.}, 
mathematicians have
generalized 
the classical McKay Correspondence  \cite{threedim,ito-reid,batyrev}
to geometry.

The geometric McKay Correspondence in 2-dimensions
actually identifies the cohomology ring of
$\mt$ and the representation ring of $\Gamma$ not only as vector spaces
but as rings.
Given a finite subgroup $\Gamma\subset SU(2)$, the intersection matrix
of the irreducible components of the exceptional divisor of the
resolved manifold is given by
the negative of the Cartan matrix of the associated Dynkin diagram
which is specified by the algebraic McKay Correspondence.  Hence,
there exists 
an equivalence between the tensor product
decompositions of conjugacy classes and 
intersection pairings of homology classes.  Indeed in
\cite{nakajima}, Ito and Nakajima prove that for all $\Gamma\subset SU(2)$
and for  abelian
$\Gamma\subset SU(3)$, the Groethendieck (cohomology) ring
of $\mt$ is isomorphic as a $\Z$-module 
to the representation ring of $\Gamma$ and that the intersection
pairing on  its dual, the Groethendieck
group of coherent sheaves on $\pi^{-1} (0)$, can be expressed as the
Clebsch-Gordan coefficients.  Furthermore, string theory analysis also
predicts a similar relation between the two ring structures
\cite{Dijkgraaf}.

The geometric McKay Correspondence can thus be stated as
\begin{conjecture}[\sc Geometric McKay Correspondence] \label{gmc} Let
$\Gamma,M,$
and $\mt$ be as in Conjecture~\ref{euler}.  Then, there exist bijections
\barrayn
\mbox{\fbox{\rule[-3mm]{0mm}{9mm}  Basis of $H^* (\mt,\Z)$}} \ \
	&\longleftrightarrow&\ \  
	\mbox{\fbox{\rule[-3mm]{0mm}{9mm} $\{$Irreducible
Representations of  
$\Gamma \}$}}\\
\mbox{\fbox{\rule[-3mm]{0mm}{9mm}  Basis of $ H_* (\mt,\Z)$}} \ \
	&\longleftrightarrow&\ \  
	\mbox{\fbox{\rule[-3mm]{0mm}{9mm} $\{$Conjugacy Classes of
$\Gamma \}$}}\ , 
\earrayn
and there is an identification between the two ring structures.
\end{conjecture}

\subsubsection{Question of Ito and Reid and Chiral Ring}
In \cite{ito-reid}, Ito and Reid raised the question whether the
cohomology ring\footnote{Henceforth, $\dim M=n$
 is not restricted to 2.}
 $H^*(\widetilde{M})$ is generated by
$H^2(\widetilde{M})$.  In this subsection, we rephrase the question
in terms of $N=2$ SCFT on $M=\C^n/\Gamma$, $\Gamma\subset SU(n)$.
String theory provides a way\footnote{It is believed that string
theory somehow picks out a distinguished resolution of the orbifold,
and the following discussion pertains to such a resolution when it
exists.} of computing the cohomology of the resolved manifold $\mt$.
Let us briefly review the method for the present case \cite{orbifold}:  

The cohomology of $\mt$ consists of those elements of $H^*(\C^n)$ that
survive the projection under $\Gamma$ and new classes arising from the
blow-ups.  In this case, $H^0(\C^n)$ is a set of all constant functions
on $\C^n$ and survives\footnote{This cohomology class should 
correspond to the trivial
representation in the McKay correspondence.} 
the projection, while all other cohomology classes vanish.  Hence, all
other non-trivial elements of $H^*(\mt)$ arise from the blow-up
process; in string theory language, they correspond to the
twisted chiral primary operators, which are not necessarily all marginal.
In the $N=2$ SCFT of non-linear sigma-model on a compact CY manifold, 
the $U(1)$
spectral flow 
identifies the chiral ring of the SCFT with the cohomology ring of the
manifold, modulo
quantum corrections.  
For non-compact cases, by
considering a topological non-linear $\sigma$-model, 
the $A$-model chiral ring matches the cohomology ring and
the blow-ups still correspond to the twisted sectors.

 An $N=2$ non-linear sigma model on  a CY $n$-fold $X$
 has two topological
twists called the $A$ and $B$-models, of which the ``BRST'' non-trivial
observables \cite{tqft} encode the information about the K\"{a}hler and 
complex structures of $X$, respectively.
The correlation functions of the $A$-model receive instanton corrections
whereas the classical computations of the $B$-model give exact quantum
answers.  The most efficient way of computing the $A$-model
correlation functions is to map the theory to a $B$-model on another
manifold $Y$ which is a mirror\footnote{Mirror symmetry has been
intensely studied by both mathematicians and physicists for the past
decade, leading to many powerful tools in enumerative geometry.  A detailed
discussion of mirror symmetry is beyond the scope of this paper, and we
refer the reader to \cite{mirror} for introductions to the subject and
for references.} of $X$ \cite{mirror}.  Then, the classical
computation of the $B$-model on $Y$ yields the full quantum answer
for the $A$-model on $X$. 

In this paper, we are interested in K\"{a}hler resolutions of the Gorenstein
singularities and, hence, in the $A$-model whose chiral ring is a quantum 
deformation of the classical cohomology ring.
Since all non-trivial elements of the cohomology
ring, except for $H^0$, arise from the twisted sector or blow-up
contributions, we have
the following reformulation of the Geometric McKay
Correspondence which is well-established in string theory:
\begin{proposition}[{\sc String Theory McKay
Correspondence}]\label{conjecture1} 
Let $\Gamma$ be a
discrete subgroup of $SU(n)$ such that the Gorenstein singularities of
$M=\C^n/\Gamma$ has a crepant resolution $\pi:\mt\ra M$.  Then, there
exists a following bijection between the cohomology and $A$-model data:
\beq
	\mbox{\fbox{\rule[-3mm]{0mm}{9mm} Basis of $\bigoplus\limits_{i>0}
H^i (\mt)$} \ \ 
	$\longleftrightarrow$\ \  
	\fbox{\rule[-3mm]{0mm}{9mm} $\{ \mbox{Twisted Chiral Primary
	Operators}\}$}}\ ,
\eeq
or equivalently, by the Geometric McKay Correspondence,

\beq
	\mbox{\fbox{\rule[-3mm]{0mm}{9mm} \{Conjugacy classes of
	$\Gamma$\}}\ \ 
	$\longleftrightarrow$\ \  
	\fbox{\rule[-3mm]{0mm}{9mm} \{Twisted Elements of the Chiral
	Ring\}}}\ . 
\eeq
\end{proposition}
Thus, since all $H^i, i>0$ arise from the twisted chiral primary but
not necessarily marginal fields and since the marginal operators
correspond to $H^2$,
we can now reformulate the question of whether $H^2$ generates $H^*$ as
follows: 
\begin{quote}
{\it Do the marginal twisted
 chiral primary fields generate the entire twisted  chiral ring?}
\end{quote}
\noindent
This kind of string theory resolution of orbifold singularities
is Arrow III in Figure~\ref{fig:mother}.  In \sref{sec:CFT}, we will
see how a conformal field theory description of
the singular limit of these string theories naturally 
allows us to link geometry to representation theory.  In this way, we
hint why McKay correspondence and the discoveries of \cite{CIZ} 
are not mere happy flukes of nature, as it will become clearer as
we proceed.

\subsection{(I, IV) McKay Correspondence and WZW} \label{subsec:Mckay-WZW}
When we calculate the partition function for the WZW
model with its energy momentum tensor associated to an algebra
$\widehat{g_k}$ 
of level $k$, it will be of the form\footnote{we henceforth
use the notation in \cite{CFT}}:
\[
Z(\tau) = \sum_{\widehat{\lambda},\widehat{\xi} \in P^{(k)}_+}
	\chi_{\widehat{\lambda}}(\tau)
	{\cal M}_{\widehat{\lambda},\widehat{\xi}}
	\overline{\chi_{\widehat{\xi}}}(\overline{\tau})
\]
where $P^{(k)}_+$ is the set of dominant weights and 
$\chi_{\widehat{\lambda}}$
is the affine character of $\widehat{g_k}$. The matrix ${\cal M}$ gives the
multiplicity of the highest weight modules in the decomposition of the
Hilbert space and is usually referred to as the {\it mass
matrix}. Therefore 
the problem of classifying the modular invariant 
partition functions of WZW models is
essentially that of the integrable characters $\chi$ of affine Lie algebras.

In the case of $\widehat{g_k} = \widehat{SU(2)_k}$, all the modular 
invariant partition functions are classified, and they fall into an \ADE 
scheme (\cite{CFT} to \cite{Gannon2}).
In particular, they are of the form of sums over
modulus-squared of combinations of the weight $k$ Weyl-Kac character
$\chi^k_\lambda$ for $\widehat{SU(2)}$ (which is in turn
expressible in term of Jacobi theta functions), where
the level $k$ is correlated with the rank of the ADE Dynkin diagrams as 
shown in Table~\ref{table:ADE-WZW} and $\lambda$ are the eigenvalues
for the 
adjacency matrices of the \ADE Dynkin diagrams.
\begin{table}
\begin{center}
\begin{tabular}{||c|l||}  \hline
Dynkin Diagram of Modular Invariants &  Level of WZW\\ \hline
$A_n$ & $n-1$\\ \hline
$D_n$ & $2n-4$\\ \hline
$E_6$ & 10 \\ \hline
$E_7$ & 16 \\ \hline
$E_8$ & 28 \\ \hline
\end{tabular}
\caption{The $ADE$-Dynkin diagram representations of the 
modular invariants of the $\widehat{SU(2)}$ WZW. \label{table:ADE-WZW}}
\end{center}
\end{table}
Not only are the modular invariants classified by these graphs,
but some of the fusion ring algebra can be reconstructed from
the graphs.

Though still largely a mystery, the reason for this classification
can be somewhat traced to the so-called {\it fusion rules}.
In a rational conformal field theory, the fusion coefficient 
$N_{\phi_i \phi_j}^{\phi_k^*}$ is defined by

\begin{equation}
\phi_i \times \phi_j = \sum\limits_{\phi_k^*} 
{\cal N}_{\phi_i \phi_j}^{\phi_k^*}
	\phi_k^*
\label{fusion}
\end{equation}
\noindent
where $\phi_{i,j,k}$ are chiral\footnote{Chirality here means left- or 
right-handedness not chirality in the sense of $N=2$ superfields.} 
primary fields. This fusion rule provides 
such vital information as the number of independent coupling between the
fields and the multiplicity of the conjugate field $\phi_k^*$ appearing in
the operator product expansion (OPE) of $\phi_i$ and $\phi_j$.
In the case of the WZW model with the energy-momentum tensor taking values
in the algebra $\widehat{g_k}$ of level $k$, we can recall 
that the primary fields have integrable representations
$\widehat{\lambda}$ in the dominant weights of $\widehat{g_k}$, and
subsequently, (\ref{fusion}) reduces to
\[
\widehat{\lambda} \otimes \widehat{\mu} =
\bigoplus_{\widehat{\nu} \in P_+^k} 
{\cal N}_{\widehat{\lambda}\widehat{\mu}}^{\widehat{\nu}}\ \widehat{\nu}.
\]
Indeed now we see the resemblance of (\ref{fusion}) coming from
conformal field theory to (\ref{CG}) coming from finite representation
theory, hinting that there should be some underlying relation.
We can of course invert (\ref{CG}) using the properties of finite
characters, just as we can extract $\cal{N}$ by using the Weyl-Kac
character formula (or by the Verlinde equations).

Conformal field theorists, inspired by the \ADE classification
of the minimal models, have devised similar methods to treat the fusion
coefficients. It turns out that in the simplest cases the fusion
rules can be generated entirely from one special case of 
$\widehat{\lambda} = f$, the so-called fundamental representation. This
is of course in analogy to the unique (fundamental) 2-dimensional 
representation $R$ in McKay's paper. In this case, all the information
about the fusion rule is encoded in a matrix $[N]_{ij} = {\cal
N}_{fi}^j$, 
to be treated as the adjacency matrix of a finite graph. Conversely we
can define a commutative algebra to any finite graph, whose
adjacency matrix is defined to reproduce the fusion rules for
the so-called {\it graph algebra}. It turns out that in the cases of
$A_n, D_{2n}, E_6$ and $E_8$ Dynkin diagrams, 
the resulting graph algebra has an
subalgebra which reproduces the (extended)
fusion algebra of the respective \ADE $\widehat{SU(2)}$ WZW models.

From another point of view, we can study the WZW model by quotienting
it by discrete subgroups of $SU(2)$; this is analogous to the
twisted sectors in string theory where for the partition function we
sum over all states invariant under the action of the discrete 
subgroup. Of course in this case we also have an $A$-$D$-$E$-type
classification 
for the finite groups due to the McKay Correspondence, therefore
speculations have risen as to why both the discrete subgroups and
the partition functions are classified by the same graphs 
\cite{CFT,DiFrancesco}, which also reproduce the associated ring 
structures.
The reader may have noticed that this connection
is somewhat weaker than the others hitherto considered, in the sense
that the adjacency matrices do not correspond 1-1 to the fusion rules.
This subtlety will be addressed in \sref{sec:CFT} and \sref{sec:ribbons}.

Indeed, the graph algebra construction has been extended to $\widehat{SU(3)}$
and a similar classification of the modular invariants have in fact been done
and are shown to correspond to the so-called {\it generalized Dynkin
Diagrams} \cite{CFT,Gannon1,DiFrancesco}. On the other hand,
the Clebsch-Gordan coefficients of the McKay type for the discrete subgroups
of $SU(3)$ have been recently computed in the context of studying 
D3-branes on orbifold singularities \cite{he}. It was noted that the
adjacency 
graphs drawn in the two different cases are in some form of correspondence
and was conjectured that this relationship might extend
to $\widehat{SU(n)_k}$ model for $n$ other than 2 and 3 as well. It is
hoped that this problem may be attacked by going through the other
arrows.

We have now elucidated arrows I and IV in \fref{fig:mother}.

\setall
\section{The Arrow V: $\sigma$-model/LG/WZW Duality}  \label{sec:CFT}
We here summarize the link V in \fref{fig:mother} for ALE spaces, as has been
established in \cite{Ooguri-Vafa}.

It is well-known that
application of catastrophe theory leads to the \ADE
classification of Landau-Ginzburg models \cite{vafa2}.  It has been
subsequently shown 
that the renormalization group fixed points of these theories
actually provide the Lagrangian formulations of
$N=2$ discrete minimal models \cite{witten}.  What is even more
surprising and beautiful is Gepner's another proposal \cite{gepner} that 
certain classes of
$N=2$ non-linear sigma-models on CY 3-folds are equivalent to 
tensor products of $N=2$ minimal models with the correct central charges and
$U(1)$ projections.  
Witten has successfully verified the claim  in
\cite{witten2}  using a gauged linear-sigma model which interpolates
between Calabi-Yau compactifications and Landau-Ginzburg orbifolds.

In a similar spirit, Ooguri and Vafa have considered  LG orbifolds\footnote{
The universality classes of  the LG models are completely specified by their
superpotentials $W$, and such a simple characterization leads to very powerful 
methods of detailed computations \cite{vafa1,Vafa-LG}.  Generalizations
of these models have many important applications in string theory, and
the OPE coefficients of topological LG theories with judiciously chosen
non-conformal deformations yield the fusion algebra of rational 
conformal field theories.
In \cite{gepner2}, Gepner has shown 
that the topological
 LG models with deformed Grassmannian superpotentials yield the fusion
algebra of the $\widehat{SU(n)}_k$ WZW, illustrating
 that much information about
 non-supersymmetric RCFT can be extracted from
their $N=2$ supersymmetric counterparts.  Gepner's superpotential could be
viewed as a particular non-conformal deformation of the
superpotential appearing in Ooguri and Vafa's model.} of
the tensor product of $SL(2,\R)/U(1)$ and $SU(2)/U(1)$
Kazama-Suzuki models\footnote{The $SL(2,\R)/U(1)$ coset model describes the
two-dimensional black hole geometry \cite{Witten-black}, while
the $SU(2)/U(1)$ Kazama-Suzuki model is just
the $N=2$ minimal model.} \cite{Kazama} and have shown that the resulting
theory describes the singular conformal field theory of the non-linear
sigma-model with the $B$-field turned off.  In particular,  they have
shown that the singularity on $A_{n-1}$ ALE space is
described by the
\beq
	\frac{\frac{SL(2)_{n+2}}{U(1)} \times \frac{SU(2)_{n-2}}{U(1)}}{\Z_n}
\eeq
orbifold model which contains the $\widehat{SU(2)}_{n-2}$ WZW theory
at level $k=n-2$.  The coset descriptions of the non-linear $\sigma$-models on 
$D$ and $E$-type ALE spaces also contain the corresponding WZW theories whose
modular invariants are characterized by the $D$ and $E$-type resolution graphs
of the ALE spaces.  The full orbifoldized Kazama-Suzuki model has
fermions as well 
as an extra Feigin-Fuchs scalar, but we will be interested only in the
WZW sector 
of the theory, for this particular sector
contains the relevant information about the discrete group
$\Gamma$ and the cohomology of $\widetilde{\C^2/\Gamma}$.
We summarize the results in \tref{table:ALE-WZW}.

\begin{table}
\begin{center}
\begin{tabular}{||c|l||}  \hline
ALE Type &  Level of WZW\\ \hline
$A_n$ & $n-1$\\ \hline
$D_n$ & $2n-4$\\ \hline
$E_6$ & 10 \\ \hline
$E_7$ & 16 \\ \hline
$E_8$ & 28 \\ \hline
\end{tabular}
\caption{The WZW subsector of the Ooguri-Vafa conformal field theory
 description of the 
 singular non-linear sigma-model 
on ALE. \label{table:ALE-WZW}}
\end{center}
\end{table}

We now assert that
many amazing \ADE-related properties of the $\widehat{SU(2)}$ WZW conformal 
field theory and the
McKay correspondence can be interpreted as consequences of the fact that the
conformal field theory description of the singularities of ALE spaces contains
the $\widehat{SU(2)}$ WZW.  That is, we argue that the WZW theory 
inherits most of the geometric information about the ALE spaces.

\subsection{Fusion Algebra, Cohomology and Representation Rings}
Comparing the Table~\ref{table:ALE-WZW} with the Table~\ref{table:ADE-WZW},
we immediately see that the graphical representations of the homology intersections
of $H_2(\widetilde{\C^2/\Gamma},\Z)$
and the modular invariants of the associated
$\widehat{SU(2)}$ WZW subsector are identical.

Let us recall 
how $\widehat{SU(2)}_k$ WZW model has been historically 
related to the finite subgroups of
$SU(2)$. Meanwhile we shall recapitulate some of the key points in 
\sref{subsec:Mckay-WZW}.
The finite subgroups $\Gamma$ of $SU(2)$ 
have two infinite and one finite series.  The Algebraic McKay Correspondence
showed that the representation ring of each finite group admits a graphical 
representation such that the two infinite series have the precise $A$ and $D$
Dynkin diagrams while the finite series has the $E_{6,7,8}$ Dynkin diagrams.
Then, it was noticed that the same Dynkin diagrams classify 
the modular invariants of the $\widehat{SU(2)}_k$ WZW model, and this 
observation was interesting but there was no {\it a priori}\/ connection 
to the representation theory of finite subgroups.  It was later 
discovered that the Dynkin diagrams also encode the $\widehat{SU(2)}_k$ WZW
fusion rules or their extended versions\footnote{See \cite{CFT} for a more
complete discussion of this point.}.  Independently of the WZW models,
the Dynkin diagrams are also well-known to 
represent the homological intersection numbers on $\widetilde{\C^2/\Gamma}$, 
which are encoded the chiral ring structure of the sigma-model when $B\neq0$.
What Ooguri and Vafa have shown us is that when the $B$-field is set to zero,
the information about the chiral ring and the
discrete subgroup $\Gamma$ do not get destroyed but get transmitted to the
orbifoldized Kazama-Suzuki model which contains the $\widehat{SU(2)}_k$ WZW.

Let us demonstrate the fusion/cohomology correspondence for the
$A$-series.  Let $C_i$ be the basis of $H^2(\widetilde{\C^2/\Z_n},\Z)$ and
$Q_{ij}$ their intersection matrix inside the $A_{n-1}$ ALE space.  The
$\widehat{SU(2)}_k$ WZW at level $k=n-2$ has $k+1$ primary fields $\phi_a, 
a=0, 1,\ldots n-2$.  Then, the fusion of the fundamental field $\phi_1$ with
other primary fields 
\beq
	\phi_1 \times \phi_a = {{\cal N}_{1a}}^b\ \phi_b
\eeq
is precisely given by the intersection matrix, i.e. ${{\cal N}_{1a}}^b=Q_{ab}$.
Now, let $N_1$ be the matrix whose components are the fusion coefficients
$(N_1)_{ab}={{\cal N}_{1a}}^b$, and define
$k-1$ matrices $N_i, i=2,\ldots,k$ recursively by the following equations
\barray
	N_1N_1 &=& N_0 +N_2 \nonumber\\
	N_1N_2 &=& N_1+N_3\nonumber\\
	N_1N_3 &=& N_2 + N_4 \ \cdots\nonumber\\ 
	N_1N_{k-1} &=& N_{k-2} + N_{k}\nonumber\\
	N_1N_k  &=& N_{k-1} \label{eq:graph-alg}
\earray
where $N_0=\mbox{Id}_{(k+1)\times (k+1)}$.  That is, multiplication by $N_1$ with
$N_j$ just lists the neighboring nodes in the $A_{k+1}$ Dynkin diagram with a
sequential labeling.  Identifying the primary fields $\phi_i$ with the matrices
$N_i$, it is easy to see that the algebra of $N_i$ generated by 
the defining equations \eqr{eq:graph-alg}
precisely reproduces the fusion algebra of the $\phi_i$ for the
$\widehat{SU(2)}_k$ WZW at level 
$k=n-2$.  This algebra is the aforementioned  graph algebra in conformal field 
theory.  The graph algebra has been known for many years, but what we
are proposing 
in this paper is that the graph algebra is a 
consequence of the fact that the WZW 
contains the information about the cohomology of the corresponding ALE space.

Furthermore, recall from \sref{sec:alg-mckay} 
that the intersection matrix is identical to
the Clebsch-Gordan coefficients $m_{ij}$, ignoring the affine node.  This fact is in
accordance with the proof of Ito and Nakajima \cite{nakajima} that the
cohomology ring of $\widetilde{\C^2/\Gamma}$ is isomorphic to the representation
ring ${\cal R}(\Gamma)$  
of $\Gamma$.  At first sight, it appears that we have managed to reproduce
only a subset of Clebsch-Gordan coefficients of ${\cal R}(\Gamma)$ 
from the cohomology or equivalently the fusion
ring.  For the $A$-series, however, we can easily find all the Clebsch-Gordan 
coefficients of the irreps of $\Z_n$ from the fusion algebra by simply
relabeling the irreps and choosing a different
 self-dual 2-dimensional representation.  This is because the algebraic
McKay correspondence produces an $A_{n-1}$ Dynkin diagram 
for any self-dual 2-dimensional representation $R$ and choosing a different
$R$ amounts to relabeling the nodes with different irreps.  
The graph algebras of the $\widehat{SU(2)}_k$
WZW theory for the $D$ and $E$-series actually lead not to the fusion
algebra of the original theory but to that of the extended theories, and
these cases require further investigations.

\begin{figure}[ht]
\centerline{\psfig{figure=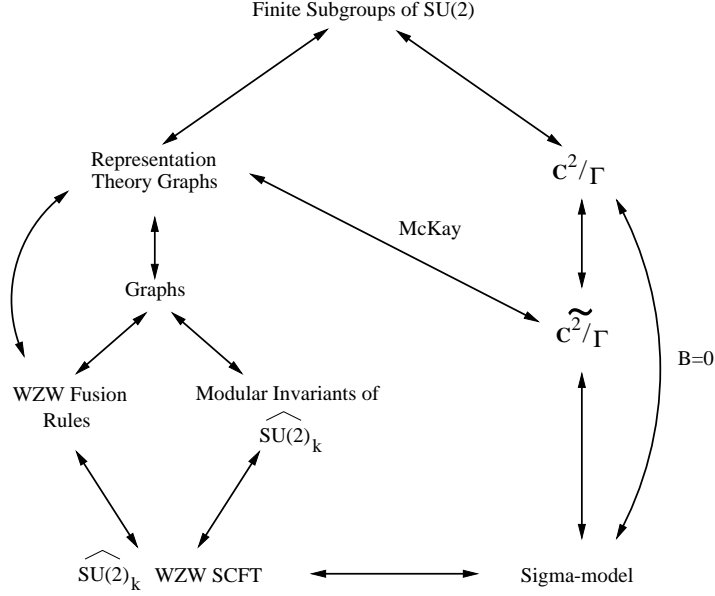,width=3.7in}}
\caption{Web of Correspondences:  \small Each finite group $\Gamma\subset SU(2)$ 
gives rise to an isolated  Gorenstein singularity as well as to its representation
ring ${\cal R}$.  The cohomology ring of the resolved manifold is
isomorphic to 
${\cal R}$.
The  $\widehat{SU(2)}_k$ WZW theory at level 
$k= \mbox{$\#$ Conjugacy classes of $\Gamma$}-2$ has a graphical
representation of its 
modular invariants and its fusion ring.  The resulting graph is
precisely the non-affine 
version of McKay's graph for $\Gamma$.  The WZW model arises as a
subsector of the  
conformal field theory description of the quotient singularity when
the $B$-field has 
been set to zero.  We further note that the three rings in the picture
are equivalent. 
\label{fig:2-dim}}
\end{figure}

String theory is thus telling us that the cohomology ring of
$\widetilde{\C^2/\Gamma}$, fusion ring of $\widehat{SU(2)}$ WZW and
the representation ring of $\Gamma$ are all equivalent.
We summarize the noted correspondences and our observations in 
Figure~\ref{fig:2-dim}.

%
\subsection{Quiver Varieties and WZW}
In this subsection, we suggest how affine Lie algebras may be arising so
naturally in the study of two-dimensional quotient spaces.

Based on the previous studies of Yang-Mills instantons on ALE spaces as in
\cite{Kronheimer},
Nakajima has introduced in \cite{Nakajima-quiver} the notion of a quiver
variety which is roughly a hyper-K\"{a}hler 
moduli space of representations of a quiver associated to a finite 
graph (We shall turn to quivers in the next section). There, he presents a beautiful
 geometric construction of representations of affine Lie algebras. 
In particular, he shows that when the graph is of the \ADE type, 
the middle cohomology of the quiver variety is isomorphic to the weight space of
integrable highest-weight representations.  A famous example of a
quiver variety with this kind of affine Lie algebra symmetry is the
moduli space of instantons over ALE spaces.
 
In a separate paper \cite{nakajima}, Nakajima also shows that the quotient space
$\C^2/\Gamma$  admits a Hilbert scheme resolution $X$ which itself can be
identified with 
a quiver variety associated with the affine Dynkin diagram of $\Gamma$.
The analysis of \cite{Nakajima-quiver} thus seems to suggest that the second 
cohomology of the resolved space $X$ is isomorphic to the weight space of
some affine Lie algebra.  
We interpret Nakajima's work as telling physicists that the
$\widehat{SU(2)}_k$ WZW has every right to be present and
carries the geometric information about the second cohomology.
Let us demonstrate our thoughts when $\Gamma = \Z_n$.  In this case,
we have $\dim H^2 = n-1$, consisting of $n-1$ $\P^1$ blow-ups in a linear
chain.  We interpret the $H^2$ basis as furnishing a
representation  
of the $\widehat{SU(2)}_k$ WZW at level $k=n-2$, as
the basis matches the primary fields
of the WZW.  This interpretation agrees with the analysis of Ooguri and 
Vafa, but we are not certain how to reproduce the result directly from
Nakajima's work.

\subsection{T-duality and Branes}
In \cite{callan,SD,Gomis,ABKS}, the $\widehat{SU(2)}_k$ WZW theory arose in
a different 
but equivalent context of brane dynamics.  As shown in
\cite{Ooguri-Vafa}, the type IIA (IIB) string theory on an
$A_{n-1}$ ALE space is $T$-dual to
the type IIB (IIA) theory in the background of $n$ $NS5$-branes.
The world-sheet description of the
near-horizon geometry of the colliding $NS5$-branes is
 in terms of the $\widehat{SU(2)}_k$ WZW, a Feigin-Fuchs boson, and
their superpartners.  More precisely, the near-horizon geometry of $n$
$NS5$-branes is given by the WZW at level $n-2$, which is consistent with the
analysis of Ooguri and Vafa.

It was  conjectured in \cite{SD}, and further generalized in
\cite{Gomis}, that the string theory on the near horizon geometry of
the $NS5$-branes is dual to the decoupled theory on the world-volume
of the $NS5$-branes.  In this paper, our main concern has been the
singularity structure of the ALE spaces, and  we
have  thus restricted ourselves only
to the transverse directions of the $NS5$-branes in the $T$-dual
picture.

\section{Ribbons and Quivers at the Crux of Correspondences} \label{sec:ribbons}
There is a common theme in all the fields relevant to our observations
so far. In general we construct a theory and attempt to encode its
rules into 
some matrix, whether it be fusion matrices, Clebsch-Gordan coefficients, or
intersection numbers. Then we associate this matrix with some graph by
treating the former as the adjacency matrix of the latter and study the
properties of the original theory by analyzing the graphs\footnote{There is 
interesting work done to formalize to sub-factors and to investigate
the graphs generated \cite{Evans}.}.

Therefore there appears to be two steps in our program: firstly, we
need to study the commonalities in the minimal set of axioms in these
different fields, and secondly, we need to encode information
afforded by these axioms by certain graphical representations. It turns out
that there has been some work done in both of these steps, the first exemplified by
the so-called ribbon categories and the second, quiver categories.

\subsection{Ribbon Categories as Modular Tensor Categories}
Prominent work in the first step has been done by A. Kirillov
\cite{Kirillov} and we shall adhere to his notations. We are interested in
monoidal additive categories, in particular, we need the following:
\begin{definition}
A {\bf ribbon category} is an additive category $\cal C$ with the
following additional structures:
\begin{itemize}
\item BRAIDING: A bifunctor $\otimes:\cal C\times \cal C\to \cal C$
 along with 
 functorial associativity and commutativity isomorphisms for objects $V$
 and $W$:
\[\begin{array}{c}
a_{V_1,V_2,V_3}:(V_1\otimes V_2)\otimes V_3\to V_1\otimes(V_2\otimes V_3),\\
\check R_{V,W}:V\otimes W\to W\otimes V;
\end{array}\]
\item MONOIDALITY: A unit object 
${\bf 1} \in \mbox{Obj }\cal C$ along with isomorphisms
${\bf 1}\otimes V\to V, V\otimes {\bf 1}\to V$;
\item RIGIDITY of duals: for every object $V$ we have a (left) dual
$V^*$ and homomorphisms
\[\begin{array}{c}
e_V: V^*\otimes V\to {\bf 1}, \\
i_V:{\bf 1}\to V\otimes V^*;
\end{array}\]
\item BALANCING: functorial isomorphisms $\theta_V:V\to V$, 
        satisfying the compatibility condition
\[
\theta_{V\otimes W}= \check R_{W,V} \check R_{V, W}(\theta_V\otimes
\theta_W).
\]
\end{itemize}
\end{definition}
Of course we see that all the relevant rings in
Figure~\ref{fig:mother} fall under
such a category.  Namely, we see that the representation rings of
finite groups, chiral rings of non-linear $\sigma$-models, Groethendieck
rings of exceptional divisors or fusion rings of WZW, together with their
associated tensor products, are all different realizations of
a ribbon category \footnote{Of course they may possess additional
structures, e.g., these rings are all finite. We shall later see how
finiteness becomes an important constraint when going to step two.}.
This fact is perhaps obvious from the point of view of orbifold string
theory, in which the fusion ring naturally satisfies the
representation algebra of the finite group and the WZW arises as a
singular limit of the vanishing $B$-field.
The ingredients of each of these rings, respectively the irreps, chiral
operators and cohomology elements, thus manifest as the objects in $\cal C$.
Moreover, the arrows of \fref{fig:mother}, loosely speaking, 
become functors among these various representations of $\cal C$ 
whereby making our central diagram a (meta)graph associated to $\cal C$.
What this means is that as far as the ribbon category is concerned,
all of these theories discussed so far are axiomatically identical. 
Hence indeed any underlying correspondences will be natural.

What if we impose further constraints on $\cal C$?
\begin{definition} We define $\cal C$ to be {\bf semisimple} if
\begin{itemize}
\item It is defined over some field $\K$ and all the spaces of
        homomorphisms are finite-dimensional vector spaces over $\K$;
\item Isomorphism classes of simple objects $X_i$ in $\cal C$
        are indexed by elements $i$ of some
        set $I$. This implies involution ${}^*:I\to I$ such that
        $X_i^*\simeq X_{i^*}$ (in particular, $0^*=0$);
\item ``Schur's Lemma'': $\hom (X_i, X_j) = \K\delta_{ij}$;
\item Complete Finite Reducibility: $\forall$ $V\in \mbox{Obj }\cal C$, 
        $V=\bigoplus\limits_{i\in I} N_i X_i,$ such that the sum is finite,
        i.e., almost all $N_i\in \Z_+$ are zero.
\end{itemize}
\end{definition}
Clearly we see that in fact our objects, whether they be WZW fields or
finite group irreps, actually live in a semisimple ribbon category.
It turns out that semisimplicity is enough to allow us to define composition 
coefficients of the ``Clebsch-Gordan'' type:
\[
X_i \oplus X_j = \bigoplus N_{ij}^k X_k,
\]
which are central to our discussion.

Let us introduce one more concept, namely the
matrix $s_{ij}$ mapping $X_i \to X_j$ represented graphically by the
simple ribbon 
tangle, i.e., a link of 2 closed directed cycles of maps from $X_i$ and $X_j$
respectively into themselves. The remarkable fact is that imposing that
\begin{itemize}
\item $s_{ij}$ be invertible and that
\item $\cal C$ have only a finite number of simple objects
        (i.e., the set $I$ introduced above is finite) 
\end{itemize}
naturally gives rise to modular properties. We define such semisimple ribbon
category equipped with these two more axioms as a {\bf Modular Tensor
Category}. 
If we define the matrix $t_{ij} = \delta_{ij} \theta_i$ with $\theta_i$ being
the functorial isomorphism introduced in the balancing axiom for $\cal C$, the
a key result is the following  \cite{Kirillov}:
\begin{theorem}
In the modular tensor category $\cal C$, the matrices $s$ and $t$ generate
precisely the modular group $SL(2,\Z)$.
\end{theorem}
Kirillov remarks in \cite{Kirillov} that it might seem mysterious that modular
properties automatically arise in the study of tensor categories and argues
in two ways why this may be so. Firstly, a projective action of $SL(2,\Z)$ 
may be defined for certain objects in $\cal C$. This is essentially
the construction 
of Moore and Seiberg \cite{MS} when they have found new modular invariants for
WZW, showing how WZW primary operators are objects in $\cal C$.
Secondly, he points out that geometrically one can associate a
topological quantum 
field theory (TQFT) to each tensor category, whereby the mapping
class group of the 
Riemann surface associated to the TQFT gives rise to the modular group.
If the theories in \fref{fig:mother} are indeed providing different
but equivalent
realizations of $\cal C$, we may be able to trace the 
origin of the $SL(2,\Z)$ action on the category to the WZW
modular invariant partition functions.  That is,  it seems that 
in two dimensions the \ADE scheme, which also arises in other
representations of $\cal C$, naturally classifies some kind of 
modular invariants.  In a generic realization of the modular tensor
category, it may be difficult to identify such modular invariants, but
they are easily identified as the invariant partition functions 
in the WZW theories.

\subsection{Quiver Categories}
We now move onto the second step. Axiomatic studies of the encoding procedure 
(at least a version thereof)
have been done even before McKay's result. In fact, in 1972, Gabriel has
noticed that categorical studies of quivers lead to $A$-$D$-$E$-type
classifications \cite{Gabriel}.
\begin{definition}
We define the {\bf quiver category} ${\cal L}(\Gamma,\Lambda)$,
for a finite connected graph $\Gamma$ with orientation $\Lambda$, vertices
$\Gamma_0$ and edges $\Gamma_1$ as follows:
The objects in this category are any
collection $(V,f)$ of spaces $V_{\alpha}, \alpha \in \Gamma_0$ and mappings 
$f_{l}, l \in \Gamma_1$. The morphisms are 
$\phi : (V,f) \rightarrow (V',f') $
a collection of linear mappings $\phi_{\alpha}  : V_{\alpha}
\rightarrow V'_{\alpha}$ compatible with $f$ by 
$\phi_{e(l)}f_l = f'_{l}\phi_{b(l)}$ where $b(l)$ and $e(l)$ are the beginning
and the ending nodes of the directed edge $l$.
\end{definition}
Finally we define decomposability in the usual sense that
\begin{definition}
The object $(V,f)$ is {\bf indecomposable} iff there do not exist objects
$(V_1,f_1), (V_2,f_2) \in {\cal L}(\Gamma,\Lambda)$ such that
$V = V_1 \oplus V_2$ and $f = f_1 \oplus f_2$.
\end{definition}
Under these premises we have the remarkable result:
\begin{theorem}[{\sc Gabriel-Tits}] The graph
$\Gamma$ in ${\cal L}(\Gamma,\Lambda)$ coincides with one of the
graphs $A_n,D_n,E_{6,7,8}$, if and only if there are only finitely many
non-isomorphic indecomposable objects in the quiver category.
\end{theorem}
By this result, we can argue that the theories, which we have seen
to be different representations of the ribbon category $\cal C$ and
which all
have \ADE classifications in two dimensions, each must in fact be
realizable as a finite quiver category $\cal L$ in dimension
two. Conversely, the finite quiver category has representations as
these theories in 2-dimensions. To formalize, we state
\begin{proposition}
In two dimensions, finite group representation ring, 
WZW fusion ring, Gorenstein
cohomology ring, and non-linear $\sigma$-model chiral ring, as
representations of a ribbon category $\cal C$, can be mapped to a
finite quiver category $\cal C$. In particular the ``Clebsch-Gordan''
coefficients ${\cal N}_{ij}^k$ of $\cal C$ realize as 
adjacency matrices of graphs in $\cal L$
\footnote{Here the graphs are \ADE Dynkin diagrams. For higher
dimension we propose that there still is a mapping, though perhaps not
to a {\it finite} quiver category.}.
\end{proposition}
Now $\cal L$ has recently been given a concrete realization 
by the work of Douglas and Moore \ \cite{Douglas}, in the
context of investigating string theory on orbifolds. The objects in the
quiver category have found representations in the resulting ${\cal N}=2$
Super Yang-Mills theory. The modules $V$ (nodes) manifest themselves
as gauge groups
arising from the vector multiplet and the mappings $f$ (edges which in 
this case are really bidirectional arrows), as bifundamental matter.
This is the arrow from graph theory to string orbifold theory in the
center of \fref{fig:mother}.
Therefore it is not surprising that an \ADE type of result in encoding
the physical content of the theory has been
obtained. Furthermore, attempts at brane configurations to construct these 
theories are well under way (e.g. \cite{Kapustin}).

Now, what makes \ADE and two dimensions special?
A proof of the theorem due to Tits \cite{Gabriel} 
rests on the fact that the
problem can essentially be reduced to a Diophantine inequality in the
number of 
nodes and edges of $\Gamma$, of the general type:
\[
\sum\limits_i\frac{1}{p_i} \ge c
\]
where $c$ is some constant and $\left\{p_i\right\}$ is a set of integers
characterizing the problem at hand.
This inequality has a long history in mathematics \cite{Humphereys}.
In our context, we recall that the uniqueness of the five perfect solids
in $\R^3$ (and hence the discrete subgroups of $SU(2)$) 
relies essentially on the equation $1/p + 1/q \ge 1/2$ having only
5 pairs of integer solutions. Moreover we recall that Dynkin's classification
theorem of the simple Lie algebras depended on integer solutions of
$1/p + 1/q + 1/r \ge 1$.

Since Gabriel's theorem is so restrictive, extensions thereto have been done
to relax certain assumptions (e.g., see \cite{Nazarova}). This will
hopefully give us give more graphs and in particular those appearing
in finite group, WZW, orbifold theories or non-linear $\sigma$-models
at higher dimensions.
A vital step in the proof is that a certain quadratic form over the $\Q$-module 
of indices on the nodes (effectively the Dynkin labels) 
must be positive-definite.
It was noted that if this condition is relaxed to positive semi-definity, then
$\Gamma$ would include the affine cases 
$\widehat{A},\widehat{D},\widehat{E}$ as well.
Indeed we hope that further relaxations of the condition would admit more graphs,
in particular those drawn for the $SU(3)$ subgroups.
This inclusion on the one hand would relate quiver graphs
to Gorenstein singularities in dimension three due to the 
link to string orbifolds\footnote{In this case we get ${\cal N} = 1$ 
Super-Yang-Mills theory in 4 dimension.} and on the other 
hand to the WZW graph algebras by the conjecture of Hanany and He \cite{he}.
Works in this direction are under way. 
It has been recently suggested that since the discrete subgroups 
of SU(4,5,6,7) have also been classified \cite{Gannon2}, graphs for
these could be 
constructed and possibly be matched to the modular invariants corresponding
to $\widehat{SU(n)}$ for $n=4,..,7$ respectively. Moreover, proposals
for unified 
schemes for the modular invariants by considering orbifolds by abelian $\Gamma$
in SU(2,3,..,6) have been made in \cite{AA}.

Let us summarize what we have found. We see that the representation ring
of finite groups with its associated $(\otimes,\oplus)$, the chiral ring of
nonlinear $\sigma$-model with its $(\otimes, \oplus)$, the fusion ring of
the WZW model with its $(\times,\oplus)$ and the Groethendieck ring of resolved
Gorenstein singularities with it $(\otimes,\oplus)$ manifest
themselves as
different realizations of a semisimple ribbon category $\cal C$. 
Furthermore, the requirement of finiteness and an
invertible $s$-matrix makes $\cal C$ into a  modular tensor category. The
\ADE schemes in two dimensions, 
if they arise in one representation of $\cal C$, might naturally
appear in another. Furthermore, the quiver category $\cal L$ has a physical
realization as bifundamentals and gauge groups of SUSY Yang-Mills theories. The
mapping of the Clebsch-Gordan coefficients in $\cal C$ to the quivers in
$\cal L$ is therefore a natural origin for the graphical representations of the
diverse theories that are objects in $\cal C$.

\setall
\section{Conjectures}\label{sec:conj}
\begin{figure}[ht]
\centerline{\psfig{figure=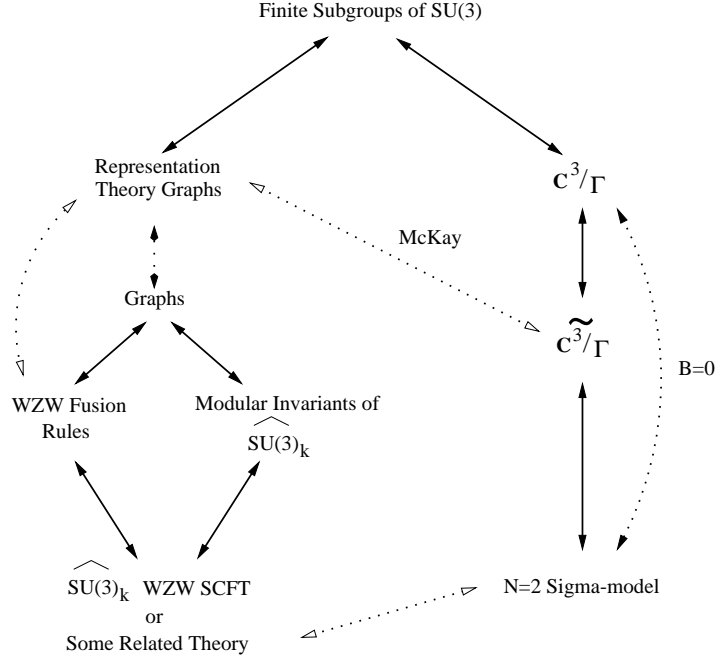,width=3.7in}}
\caption{Web of Conjectures: Recently, the graphs from the
representation theory 
side were constructed and were noted to resemble
those on WZW $\widehat{SU(3)}_k$ side \cite{he}.  The solid lines have
been sufficiently 
well-established
while the dotted lines are either conjectural or ill-defined. 
\label{fig:3-dim}}
\end{figure}
We have seen  that there exists a remarkably coherent picture of
inter-relations  
in two dimensions among many different branches of mathematics and physics.
The organizing principle appears to be the mathematical 
theory of quivers and ribbon category, while the crucial bridge
between mathematics 
and physics is the conformal field theory description of the Gorenstein
singularities provided by the orbifoldized coset construction.

Surprisingly, similar features have been noted in three dimensions.
The Clebsch-Gordan coefficients for the tensor product of
irreducible representations for all discrete subgroups of $SU(3)$ were
computed 
in \cite{he,Muto},
and a possible correspondence was noted, and conjectured for 
$n\geq 3$,
between the resulting Dynkin-like diagrams and the graphic representations of 
the fusion rules and modular invariants of $\widehat{SU(3)}_k$
WZW models.  Furthermore, 
as discussed previously, the Geometric McKay Correspondence between
the representation ring of the abelian
discrete subgroups $\Gamma\subset SU(3)$
and the cohomology ring of $\widetilde{\C^3/\Gamma}$ has been proved in
\cite{nakajima}. 
Hence, the situation in 3-dimensions as seen in \fref{fig:3-dim}
closely resembles that in 2-dimensions.

Now, one naturally inquires:

\begin{quotation}
\it
Are there graphical representations of the fusion rules and modular
invariants of the 
$\widehat{SU(n)}_k$ WZW model or some related theory
that contain the Clebsch-Gordan 
coefficients
for the representations of $\Gamma\subset SU(n)$?  And, in turn, are the 
Clebsch-Gordan coefficients related to  the (co)-homological 
intersections on the resolved geometry $\widetilde{\C^n/\Gamma}$ that
are contained in the chiral ring of the $N=2 \ \sigma$-model on $\C^n/\Gamma$ 
with a non-vanishing $B$-field?
{\bf Most importantly, what do 
these correspondences tell us about the two conformal field theories and their
singular limits?}
\end{quotation}
\noindent
As physicists, we believe that the
McKay correspondence and the classification of certain modular invariants in
terms of finite subgroups are
consequences of orbifolding and of some underlying 
quantum equivalence of the associated conformal field theories.

We thus believe that a picture similar to that seen in this paper for 2-dimensions
persists in higher dimensions and 
 conjecture that there exists a conformal field theory description of the
Gorenstein singularities in higher dimensions.
If such a theory can be found, then it would explain the observation made
in \cite{he} of the resemblance of the graphical representations of
the representation 
ring of the finite subgroups of $SU(3)$ and the modular invariants of the 
$\widehat{SU(3)}_k$ WZW.  We have checked that the correspondence, if
any, between the 
finite subgroups of $SU(3)$ and the $\widehat{SU(3)}_k$ WZW theory is
not one-to-one. 
For example, the number of primary fields generically
does not match the number of conjugacy classes
of the discrete subgroups.  It has been observed in \cite{he},
however, that some of 
the representation graphs appear to be subgraphs of the graphs encoding
the modular invariants.
We hope that the present paper serves as a motivation for finding the 
correct conformal field
theory description in three dimensions which would tell us how to
``project'' the modular invariant graphs to retrieve the 
representation graphs of the finite graphs. 

Based on the above discussions, we summarize our speculations,
relating geometry, 
generalizations of the \ADE classifications, 
representation theory, and string theory  in \fref{fig:3-dim}.  

\subsection{Relevance of Toric Geometry}
It is interesting to note that the toric resolution of
certain Gorenstein singularities also naturally admits graphical
representations of fans.  In fact, the exceptional divisors in 
the Geometric McKay Correspondence
for $\Gamma=\Z_n\subset SU(2)$ in 2-dimensions can be easily seen as
the vertices of new cones in the toric resolution, and these vertices
precisely form the $A_{n-1}$ Dynkin digram.   Thus, at least for the
abelian case in 2-dimensions, the McKay correspondence and the
classification of 
$\widehat{SU(2)}$ modular invariants seem to be most naturally
connected to geometry as toric diagrams of the resolved manifolds
$\widetilde{C^2/\Gamma}$.

Surprisingly---perhaps not so much so in retrospect---we 
have noticed a similar pattern in 3-dimensions.  That is, the toric
resolution diagrams of $\C^3/\Z_n\times \Z_n$ singularities reproduce
the graphs that classify the ${\cal A}$-type modular invariants of the
$\widehat{SU(3)}_k$ WZW models.  For which $k$?  It has been
previously observed in \cite{lattice} that there seems to be a
correspondence, up to some truncation, between the subgroups 
$\Z_n\times \Z_n\subset SU(3)$ and the ${\cal A}$-type 
$\widehat{SU(3)}_{n-1}$ modular invariants, which do appear as
subgraphs of the $\widetilde{\frac{\C^3}{\Z_n\times \Z_n}}$ toric diagrams.
On the other hand, a precise formulation of the correspondence with
geometry and the conformal field theory description of  Gorenstein
singularities still remains as an unsolved problem and will be
presented elsewhere \cite{Song}.

\section{Conclusion} \label{sec:conclusion}
Inspired by the ubiquity of \ADE classification and prompted
by an observation of a mysterious 
relation between finite groups and WZW models,
we have proposed a possible unifying scheme.
Complex and intricate
webs of connections have been presented, the particulars of which
have either been hinted at
by collective works in the past few decades in mathematics and physics
or are conjectured to exist by arguments in this paper.
These webs  include the 
McKay correspondences of various types 
as special cases and relate such seemingly disparate subjects as
finite group representation theory, graph theory, string orbifold
theory and sigma models, as well as
conformal field theory descriptions of Gorenstein singularities.
We note that the integrability of the theories that we are 
considering may play a role in understanding the deeper connections.

This paper catalogs many observations which have been
put forth in the mathematics and physics literature and presents
them from a unified perspective. 
Many existing results and conjectures have been phrased under a new light.
We can summarize the contents of this paper as follows:
\begin{enumerate}
\item In two dimensions, all of the
	correspondences mysteriously fall into an \ADE type.
	We have provided, via \fref{fig:mother},
	a possible setting how these mysteries might arise
	naturally. Moreover, we have pointed out how axiomatic works
	done by 
	category theorists may demystify some of these links.  Namely,
	we have noted that the relevant rings of the theories can be
	mapped to 
	the quiver category. 
\item   We have also discussed the possible role played by the modular
 	tensor category in our picture, in which the modular invariants
	arise very naturally.  Together with the study of the
	 quiver category and
 	quiver variety, the ribbon category seems to provide the
 	reasons for the emergence of affine Lie algebra symmetry and
 	the \ADE classification of the modular invariants.
\item	We propose the validity of our program to higher dimensions, 
	where the picture is far less clear since
	there are no \ADE schemes, though some hints of generalized graphs
	have appeared.
\item  There are three standing conjectures:
\begin{itemize}\item  We propose that there exists a conformal field theory
description of the Gorenstein singularities in dimensions greater than
two.
\item  As noted in \cite{he}, we conjecture that the modular
invariants and the fusion rings of the $\widehat{SU(n)}, n>2$ WZW,
or their generalizations, may be related to the discrete
subgroups of the $SU(n)$.
\item  Then, there is the mathematicians' conjecture that there exits
a McKay correspondence between the cohomology ring
$H^*(\widetilde{C^n/\Gamma},\Z)$ and the representation ring of
$\Gamma$, for finite subgroup $\Gamma\subset SU(n)$.
\end{itemize}
We have combined these conjectures into a web so that proving one of
them would help proving the others.
\end{enumerate}
We hope
that \fref{fig:mother} essentially commutes and that the standing
conjectures 
represented by certain arrows therewithin may be solved by investigating
the other arrows.
In this way,
physics may provide us with a possible method of attack and
explanation for 
McKay's correspondence and many other related issues, and likewise
mathematical structures may help to clarify and rigorize some
observations made from string theory.

It is the purpose of this writing to
inform the physics and mathematics community of a possibly new
direction of research which could harmonize ostensibly different and
diverse branches of mathematics and physics into a unified picture.

\vspace{0.5cm}
\noindent
{\bf Acknowledgments}

\vspace{.3cm}
\noindent
We would like to express our sincere gratitude to A. Hanany and Y.S. Song 
for reviewing our paper, B. Feng for valuable discussions, and
G. Tian for his patience and explaining the mathematical subtleties of ALE spaces.
We are especially grateful to C. Vafa for helpful comments
and for correcting our mistake on the preliminary version of the
paper.  We also thank H. Nakajima for clarifying his work.

Y.H.H. would like to express his thanks to I. Savonije, M. Spradlin 
and the Schmidts and above all
{\it Ad Catharinae Sanctae Alexandria et Ad Maiorem Dei Gloriam.}
J.S.S. would like to thank A.M.W. for inspiration.
And indeed we are eternally indebted to our parents for their ever-loving
support.


\end{document}